\documentclass[received]{JHEP3} 
\JHEPspecialurl{http://jhep.sissa.it/JOURNAL/JHEP3.tar.gz}
\usepackage{amssymb,epsfig}

\newcommand\fverb{\setbox\fverbbox=\hbox\bgroup\verb}
\newcommand\fverbdo{\egroup\medskip\noindent%
                        \fbox{\unhbox\fverbbox}\ }
\newcommand\fverbit{\egroup\item[\fbox{\unhbox\fverbbox}]}
\newbox\fverbbox

\title{Measuring the top quark mass with $m_{T2}$ at the LHC}

\author{Won Sang Cho, Kiwoon Choi, Yeong Gyun Kim and Chan Beom Park \\
        Department of Physics, KAIST, Daejeon 305--017, Korea \\
        E-mail: \email{wscho@muon.kaist.ac.kr},
               ~\email{kchoi@muon.kaist.ac.kr},
               ~\email{ygkim@muon.kaist.ac.kr},
               ~\email{lunacy@muon.kaist.ac.kr} }

\received{}
\revised{}
\accepted{}

\abstract{ We investigate the possibility to measure the top quark
mass using the collider variable $m_{T2}$  at the LHC experiment.
Monte Carlo studies of $m_{T2}$ are performed with the events
corresponding to the dilepton decays of $t\bar{t}$ produced at the
LHC with 10 $fb^{-1}$ integrated luminosity. Our analysis suggests
that the top quark mass can be determined by the $m_{T2}$ variable
alone
with a good precision at the level of 1 GeV.
}


\dedicated{}

\begin{document}
\section{Introduction}
\label{sec:introduction}
When the Large Hadron Collider (LHC) is turned on, it will serve as
a `top quark factory' \cite{atlas,cms}. The cross section for
$t\bar{t}$ pair production at the LHC is estimated to be 833 pb at the
NLO calculation \cite{ttnlo}, implying roughly 8 million $t\bar t$
pairs per year at low luminosity run (10 $fb^{-1}$/year). Such a
large number of $t\bar{t}$ events will enable us to measure the top
quark mass with high precision.

Precision measurement of the top quark mass $m_t$ is desirable in
many respects. For example, it would help to constrain the allowed
Higgs boson mass in
the Standard Model (SM).
In general, it would affect the constraints on the allowed parameter
space of various models of new physics at the TeV scale, including
the Minimal Supersymmetric Standard Model and technicolor-like
models.
The top quark mass measurement can be performed through various
methods in different channels, which have their own
advantage/disadvantage with different systematic uncertainties. In
the overall, the accuracy of $m_t$ measured at the LHC is expected
to be around 1 GeV \cite{atlas-top}.

In the SM, top quark decays mostly into a b-quark and a $W$ boson.
The $W$ boson then decays hadronically ($W \rightarrow qq'$) or
leptonically ($W\rightarrow l\nu$). Depending on the $W$ boson decay
mode, the $t\bar t$ events are divided into three channels, $i.e.,$
the dilepton channel (both $W$ bosons decay leptonically), the
lepton plus jets channel (one $W$ boson decays leptonically and the
other hadronically) and the pure hadronic channel (both $W$ bosons
decay hadronically).

The dilepton channel has a small branching fraction compared to the
lepton plus jets channel and the pure hadronic channel. It also
involves two missing neutrinos, which makes a direct event-by-event
measurement of $m_t$  not possible. However, it has a cleaner
environment, e.g. less combinatorial background and less jet energy
scale dependence, compared to other channels, therefore various
approaches for an indirect measurement of $m_t$ with dilepton
channel  have been investigated \cite{atlas-top}.

It has been shown that the collider variable $m_{T2}$ \cite{lester}
can be useful for the determination of new particle masses in the
process in which new particles are pair produced at hadron collider
and each of them decays into one invisible particle and one or more
visible particles \cite{lester,cho,ben,lester2,others}. In this
paper, we examine the possibility to determine the top quark mass
using $m_{T2}$ at the LHC experiment. For this, we perform three
Monte Carlo studies of $m_{T2}$ for the process $t\bar{t}\rightarrow
bl^+\nu\bar{b}l^-\nu$: the first which determines the endpoint value
of the $m_{T2}$ distribution  for the neutrino mass $m_\nu=0$, the
second to examine the functional dependence of $m_{T2}^{\rm max}$ on
the trial neutrino mass $\tilde{m}_\nu\neq 0$, which would determine
$m_t$ for a given value of the W boson mass $m_W$, and the third
which fits the $m_{T2}$ distribution to `template' distributions.
Our analysis suggests that the top quark mass can  be determined by
the $m_{T2}$ variable alone
with a good precision at the level of 1 GeV.

In sec.2, we briefly introduce  the $m_{T2}$ variable for the
dilepton decay of $t\bar{t}$. The results of Monte Carlo studies are
presented in sec.3, and  sec.4 is the conclusion.

\section{Transverse mass and $m_{T2}$ for top quark}
\label{sec:transversemass}

Let us consider a $t\bar{t}$ pair production and its subsequent
decay at the LHC:
\begin{eqnarray}
pp \rightarrow t\bar t\rightarrow b W^+ {\bar b} W^-.
\end{eqnarray}
In case that one of the $W$ bosons decays into leptons, one can
consider the associated transverse mass of $t\rightarrow bl\nu$,
which is defined as
\begin{eqnarray}
m_{T}^2
= m_{bl}^2 + m_\nu^2 + 2 (E_T^{bl}
E_T^{\nu}-\bold{p}_T^{bl}\cdot\bold{p}_T^{\nu}),
\end{eqnarray}
where $m_{bl}$ and $\bold{p}_T^{bl}$ denote the invariant mass and
transverse momentum of the $bl$ system, respectively, while $m_\nu$
and $\bold{p}_T^{\nu}$ are the mass and transverse momentum of the
missing neutrino, respectively. The transverse energies of the $bl$
system and neutrino are defined as
\begin{eqnarray}
E_T^{bl} \equiv \sqrt{|{\bold p_T^{bl}}|^2 + m_{bl}^2}~~~~{\rm
and}~~~~ E_T^{\nu} \equiv \sqrt{|{\bold p_T^{\nu}}|^2 + m_\nu^2}.
\end{eqnarray}

If the other W boson decays into hadrons, i.e. for the process
$t\bar{t} \rightarrow  bl\nu\bar{b}qq^\prime$,  ${\bf p}_T^\nu$ can
be read off from the total missing transverse momentum ${\bf
p}_T^{\rm miss}$. One might then construct the $m_T$ distribution
 of $t\rightarrow bl\nu$ from data, which can be used to determine
 the top quark mass $m_t$ as its shape and endpoint depend on $m_t$.
 However, to determine ${\bf p}_T^\nu$ in the process
 $t\bar{t} \rightarrow  bl\nu\bar{b}qq^\prime$, one needs to measure the full final state momenta of $\bar{t}
\rightarrow \bar{b}qq^\prime$, which by itself would determine $m_t$
in event-by-event basis. At any rate, if one uses information from
$\bar{t}\rightarrow \bar{b}qq^\prime$ to determine $m_t$, the
procedure involves more jets, which would result in  larger
uncertainties in the determined value of $m_t$.


A method to determine $m_t$ without using the hadronic decay of $W$
is to construct $m_{T2}$ for the dilepton decay
\begin{eqnarray}t\bar{t}\equiv t^{(1)}t^{(2)}\rightarrow
b^{(1)}l^{(1)}\nu^{(1)}b^{(2)}l^{(2)}\nu^{(2)}.
\end{eqnarray}
 Although each
neutrino momentum cannot be measured in this case, still the total
missing transverse momentum $\bold{p}_T^{\rm miss}=
 \bf{p}_T^{\nu (1)}+\bf{p}_T^{\nu (2)}$ can be determined experimentally.
The $m_{T2}$ variable of each event is defined as
\begin{eqnarray}
m_{T2} \equiv \min_{{\bf p}_{T}^{\nu(1)}+{\bf p}_{T}^{\nu(2)}={\bf
p}_T^{miss}} \left[ {\rm max} \{ m_T^{(1)}, m_T^{(2)} \}
\right], \label{eq:mt2def}
\end{eqnarray}
where $m_T^{(i)}$ ($i=1,2$) is the transverse mass of
$t^{(i)}\rightarrow b^{(i)}l^{(i)}\nu^{(i)}$, and the minimization
is performed over the ${\it trial}$ neutrino momenta ${\bf p}_T^{\nu
(i)}$ constrained as
\begin{eqnarray}
{\bf p}_{T}^{\nu(1)}+{\bf p}_{T}^{\nu(2)}={\bf p}_T^{miss}.
\end{eqnarray}

\begin{figure}[ht!]
\vskip 0.6cm
\begin{center}
\epsfig{figure=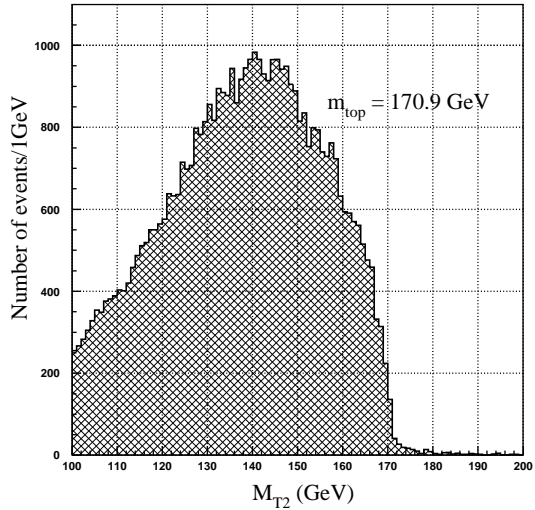,width=8cm,height=8cm,angle=0}
\end{center}
\vskip 0.4cm \caption{\it $m_{T2}$ distribution obtained from
partonic-level simulation. The input top quark mass of 170.9 GeV is
used for the simulation. One can find a sharp edge at the input top
mass, with a small tail which is mainly due to the finite top quark
decay width.} \label{fig:mt2-parton}
\end{figure}

The above definition of $m_{T2}$ indicates that $m_{T2}$ for
$m_\nu=0$ is bounded above by $m_t$ in the approximation ignoring
the decay width of top quark. One might then determine $m_t$ as
\begin{eqnarray} m_t=m_{T2}^{\rm max}(m_\nu=0) \equiv
\mbox{max}\left[\,m_{T2}(m_{bl}^{(1)},
\bold{p}_T^{bl(1)},m_{bl}^{(2)}, \bold{p}_T^{bl(2)},
{m}_\nu=0)\,\right].\end{eqnarray} In fact, because of nonzero decay
width, there can be certain amount of events which give $m_{T2}$
exceeding the physical top quark mass $m_t$. Our Monte Carlo study
suggests that such events do not spoil the sharp edge structure of
the $m_{T2}$ distribution with which one can determine $m_t$ rather
precisely. Fig. \ref{fig:mt2-parton} shows the top quark $m_{T2}$
distribution for $m_\nu=0$ obtained from a parton level Monte
Carlo simulation\footnote{For simplicity, here we switched off the
initial and final state radiations as well as the quark
fragmentation process.} using PYTHIA event generator\cite{pythia}
with an input top mass of $m_t=170.9$ GeV. One can see that $m_{T2}$
tends to zero rapidly near the input top mass with a minor but long
tail beyond the input mass which is mainly due to the nonzero top
decay width\footnote{Such a sharp edge structure of $m_{T2}$
distribution at the input mass of the mother particle can be
confirmed also in the $m_{T2}$ distribution for $W^+W^-\rightarrow
l^+\nu l^-\nu$.}.


One can consider the top quark $m_{T2}$ defined as above for
arbitrary trial neutrino mass which is {\it not} same as the true
neutrino mass. In such case, $m_{T2}$ is not only a function of the
observable kinematic variables $m_{bl}^{(i)}$ and
$\bold{p}_T^{bl(i)}$ $(i=1,2)$, but also of the trial neutrino mass.
Let $\tilde{m}_\nu$ denote the trial neutrino mass to distinguish it
from the true neutrino mass $m_\nu=0$.  The endpoint value of
$m_{T2}$ for generic $\tilde{m}_\nu$,
\begin{eqnarray} m_{T2}^{\rm
max}(\tilde{m}_\nu)=\mbox{max}\left[\,m_{T2}(m_{bl}^{(1)},
\bold{p}_T^{bl(1)},m_{bl}^{(2)}, \bold{p}_T^{bl(2)},
\tilde{m}_\nu)\,\right],\end{eqnarray} appears to be a function of
$\tilde{m}_\nu$, and its functional form provides a relation between
$m_t$, the W boson mass $m_W$, and the b quark mass $m_b$. Using the
result of Ref.\cite{cho}, one easily finds that  $m_{T2}^{max}$ as a
function of $\tilde m_\nu$ is given by
\begin{eqnarray}
m_{T2}^{\rm max}(\tilde{m}_\nu) = {m_t^2+(m_{bl}^{\rm max})^2 \over
2 m_t} + \sqrt{ \left({m_t^2-(m_{bl}^{\rm max})^2 \over 2 m_t}\right)^2+ {\tilde
m_\nu^2} }, \label{curve}
\end{eqnarray}
where
\begin{eqnarray}
(m_{bl}^{\rm max})^2 = m_b^2 +{1\over 2} (m_t^2-m_W^2-m_b^2)
+ {1\over 2} \sqrt{(m_t^2-m_W^2-m_b^2)^2-4 m_W^2 m_b^2}.
\end{eqnarray}
This analytic expression of $m_{T2}^{\rm max}(\tilde{m}_\nu)$
provides another way to determine $m_t$, i.e. one can determine
$m_t$ by fitting  $m_{T2}^{\rm max}(\tilde{m}_\nu)$ obtained from
data to this analytic expression with the known values of $m_W$ and
$m_b$.

%
%

\section{Experimental feasibility}
\label{sec:collider}


Measuring the top mass using $m_{T2}$ in real experiment will suffer
from a variety of uncertainty factors such as backgrounds, event
selection cuts, finite jet energy resolution and combinatorial
background. In order to check the feasibility of the $m_{T2}$ method
at the LHC, we have generated Monte Carlo samples of $t\bar{t}$
events by PYTHIA \cite{pythia} with the CTEQ5L parton distribution
function (PDF) \cite{pdf}. The event sample corresponds to 10 $fb^{-1}$
integrated luminosity.

The generated events have been further processed with a modified
version of fast detector simulation program PGS \cite{pgs}, which
approximate an ATLAS or CMS-like detector with reasonable
efficiencies and fake rates.
The PGS program uses a cone algorithm for jet reconstruction, with default value
of cone size $\Delta R=0.5$, where $\Delta R$ is a separation in the
azimuthal angle and pseudorapidity plane. And the b-jet tagging
efficiency $\epsilon_b$ is introduced as a function of the jet
transverse energy and pseudorapidity, with a typical value of
$\epsilon_b \sim 50\%$ in the central region for high energy jets.

In the PGS, isolated leptons (electron and muon) are identified with
some isolation cuts on the calorimeter activity around the lepton
track \cite{wiki}. For electrons, the isolation cuts are (i)
$ETISO/E_T < 0.1$, where $ETISO$ is the total transverse calorimeter
energy in a $3\times 3$ grid around the electron candidate
(excluding the candidate cell) and $E_T$ is the transverse energy of
the electron candidate, (ii) $PTISO < 5$ GeV, where $PTISO$ is the
total $p_T$ of tracks (except the electron track) with $p_T > 0.5$
GeV within a $\Delta R < 0.4$ cone around the electron candidate,
and (iii) $0.5 < EP < 1.5$, where $EP$ is the ratio of the
calorimeter cell energy to the $p_T$ of the candidate track. For
isolated muons, (i) $PTISO < 5$ GeV and (ii) $ETRAT < 0.1125$, where
$ETRAT$ is the ratio of $E_T$ in a $3\times 3$ calorimeter array
around the muon (including the muon's cell) to the $p_T$ of the
muon.

The dilepton events
are selected by requiring (A) only two isolated leptons of opposite
charge with $p_T > 25$ GeV and $|\eta|<2.5$, (B) dilepton invariant
mass with $|m_{ll}-m_Z|>5$ GeV, (C) large missing transverse energy
$E_T^{miss} > 40$ GeV, and (D) at least two b-tagged jets with
$p_T>30$ GeV and $|\eta|<3.0$.
After this selection, 5133 events are survived among the $5.5 \times 10^6$
generated $t\bar{t}$ events (in which $1.8 \times 10^5$ are the dilepton events, considering
only electrons and muons),
leading to a selection efficiency of about $2.8 \%$ for the dilepton channel signal events.

The main backgrounds might come from $Z/\gamma^*/W$ production with
additional jets, diboson events with additional jets and $b\bar{b}$
events with misidentified leptons. We have generated the main background
events using PYTHIA, ALPGEN \cite{alpgen} and AcerMC \cite{acermc},
and required the same  selection cuts as the $t\bar{t}$
dilepton events. After the cuts, it turns out that those backgrounds are
reduced to a negligible level. We will not include the background
events in our further analysis, for simplicity.

With two b-jets and two leptons in each selected event, there are
two possible combinations for $bl$ pairing. We calculated  $m_{T2}$
variable for each of the two possible $bl$ combinations, and chose
the smaller one as the final $m_{T2}$. This procedure closely
follows the idea proposed in Ref. \cite{lester2}.

\begin{figure}[ht!]
\vskip 0.6cm
\begin{center}
\epsfig{figure=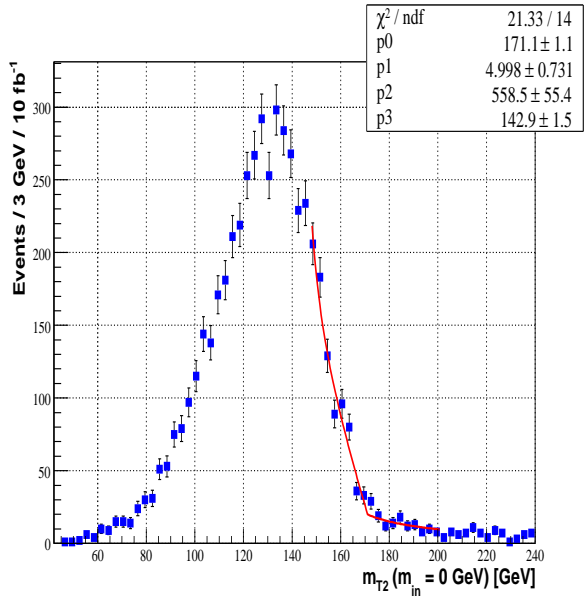,width=8cm,height=8cm}
\end{center}
\vskip 0.4cm
\caption{\it $m_{T2}$ distribution after event selection. The input value
of top quark is $m_t=170.9$ GeV. A fit to the distribution near endpoint region
is also shown, providing a fit value of $m_t=171.1\pm 1.1$ GeV.}
\label{fig:mt2}
\end{figure}
%


Fig. \ref{fig:mt2} shows the resulting $m_{T2}$ distribution for the
selected events. As anticipated, one can find an edge structure
around $m_{T2}=170$ GeV, on the distribution. We employ three
methods to precisely determine the top quark mass from the $m_{T2}$
distribution, which will be discussed in the following three
subsections.

\subsection{A fit near the end point}
\label{endfit}

Fig. \ref{fig:mt2} shows the $m_{T2}$ distribution obtained from the
selected events for the neutrino mass $m_\nu=0$. It is fitted with
an empirical function which consists of a linear function for signal
distribution and an inverse linear function for background
distribution. The fit range was chosen within $\pm {\cal O}(10)$
bins around a plausible endpoint. Such fitting of the $m_{T2}$
distribution results in
\begin{eqnarray}
m_t = 171.1\pm 1.1~ {\rm GeV},
\end{eqnarray}
which reproduces the input top quark mass of 170.9 GeV with a
precision at the level of 1 GeV.
%

To estimate possible systematic error associated with the fitting
procedure, we have repeated the fitting with two linear functions
for both signal and background distributions. The resulting top
quark mass is then given by $m_t=169.9 \pm 1.8$ GeV, showing a mass
shift of 1.2 GeV.
Systematic error from the fitting procedure might be improved by
considering a template binned likelihood fit, which will be
discussed in subsection \ref{template}.

Absolute jet energy scale also affects the determination of the top
mass. The b-jet energy scale is assumed to be known within $1\%$
accuracy. It is found that the $1\%$ variation of the jet scale
leads to a shift of the resulting top mass of 0.5 GeV.

Uncertainty due to initial state radiation (ISR) is estimated by
comparing the nominal data (with ISR switched on) to the one which
is generated while switching off ISR. The $20 \%$ of the resulting
top mass shift is found to be 0.4 GeV, which is taken as the
systematic error from ISR uncertainty \cite{atlas-top}. The same
approach to final state radiation  induces a systematic error of
0.7 GeV.

For systematic error from PDF uncertainty, 
it is found that the use of CTEQ3L (GRV94L) PDF, 
instead of the default CTEQ5L PDF,
leads to a shift of the central top mass of 0.3 (1.3) GeV, with a suitable choice of
fit range. 

\subsection{Endpoint as a function of trial neutrino mass}
\label{trial}

\begin{figure}[ht!]
\vskip 0.6cm
\begin{center}
\epsfig{figure=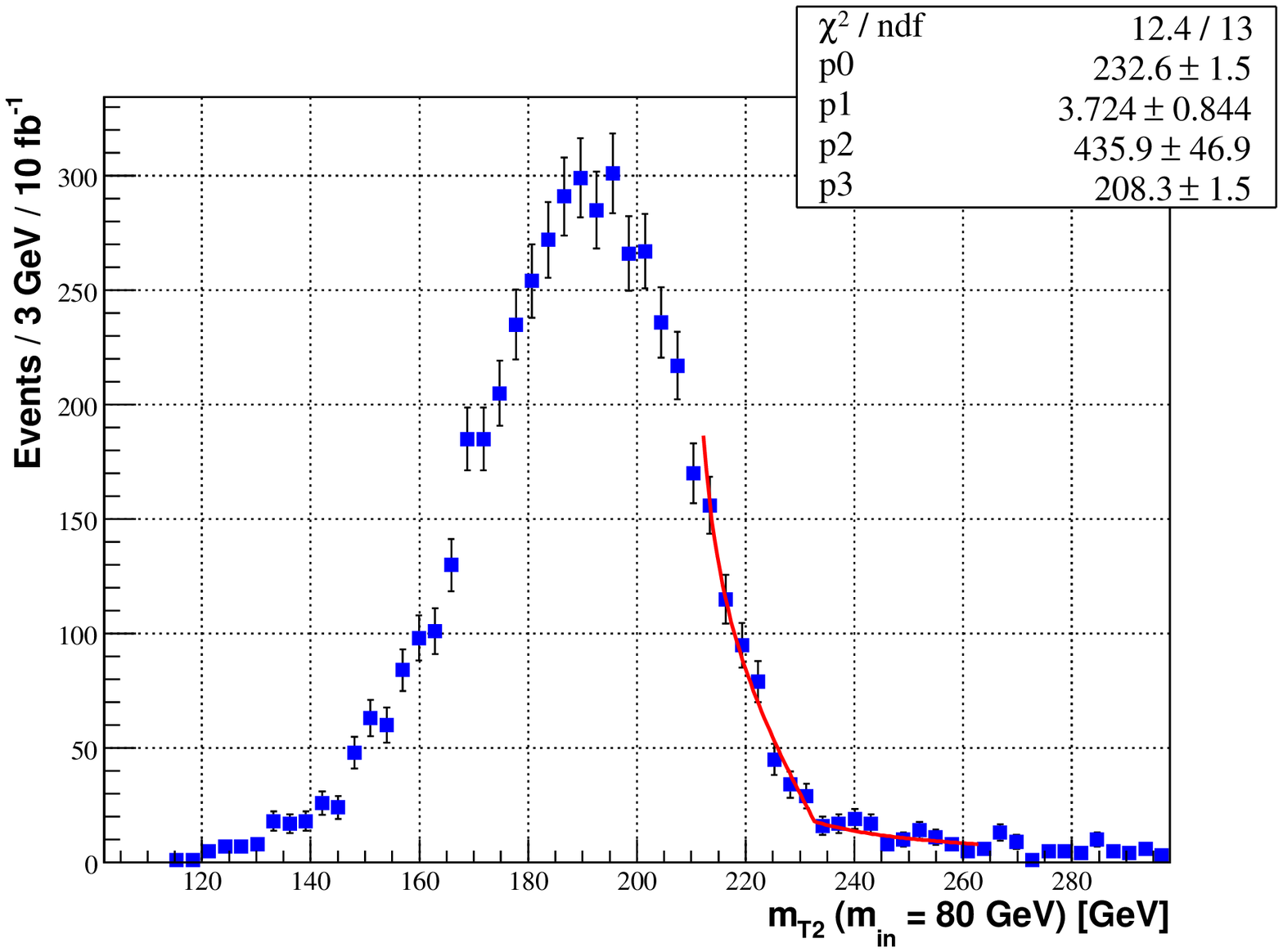,width=7cm,height=7cm}
\epsfig{figure=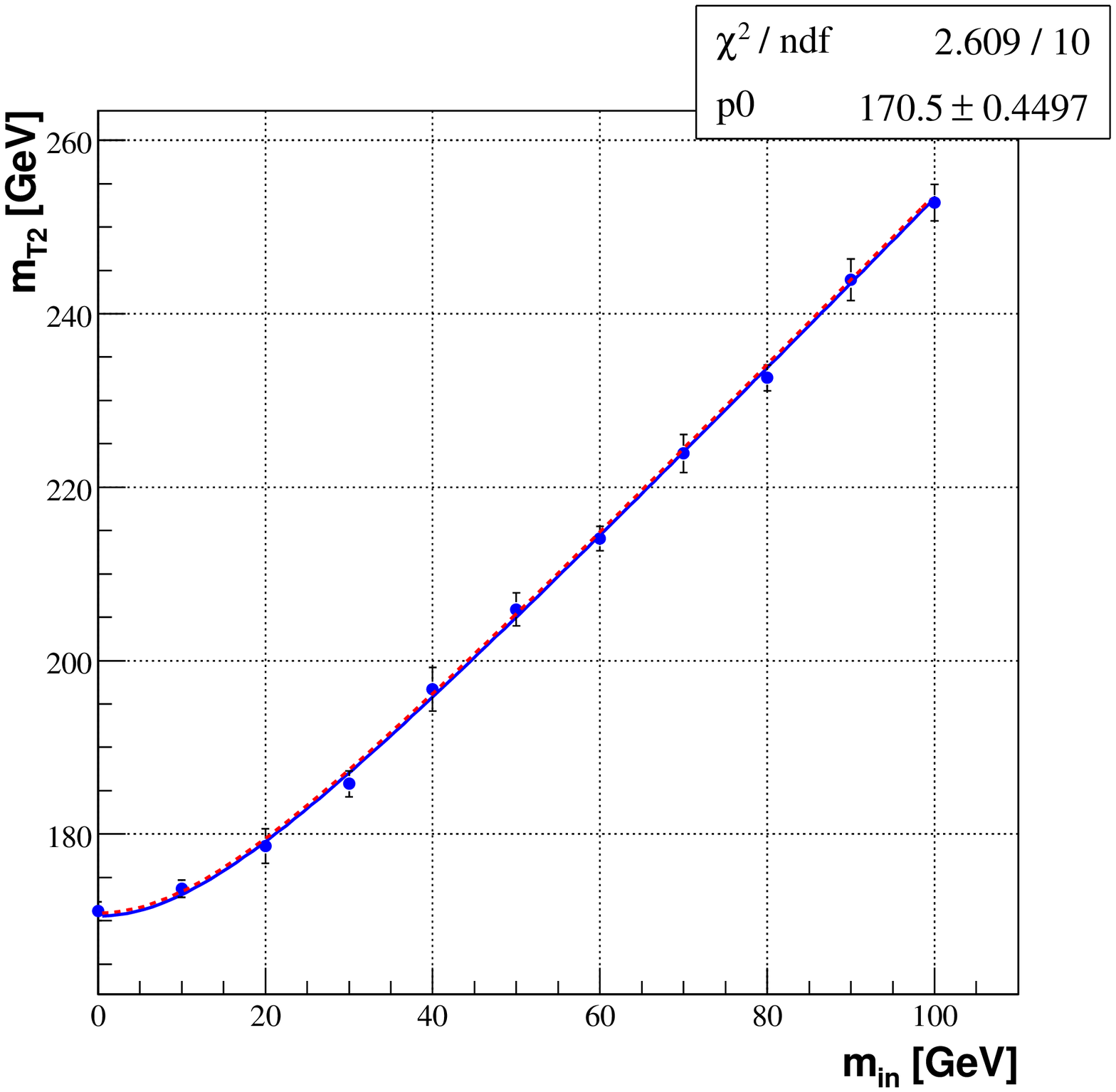,width=7cm,height=7cm}
\end{center}
\vskip 0.4cm
\caption{\it (a) An example of $m_{T2}$ distribution with a trial neutrino mass.
Here, the trial mass is set to $\tilde m_\nu=80$ GeV.
(b) The maximum of $m_{T2}$ as a function of trial neutrino mass $\tilde m_\nu$.
Also shown is the fit of the data points to theoretical curve (2.7) considering
$m_t$ as a free parameter.}
\label{fig:mt2function}
\end{figure}

As we have discussed in section \ref{sec:transversemass}, the
endpoint of $m_{T2}$ distribution can be considered as a function of
a trial neutrino mass, if we use a trial neutrino mass
$\tilde{m}_\nu\neq 0$ for the $m_{T2}$ calculation. Using the
selected dilepton decays of $t\bar{t}$, we constructed the $m_{T2}$
distributions for different  choices of $\tilde m_\nu$. Fig.
\ref{fig:mt2function}(a) shows the $m_{T2}$ distribution for $\tilde
m_\nu=80$ GeV. Here we also performed a fit to the $m_{T2}$
distribution with a linear function for signal and an inverse linear
function for background. The maximum of $m_{T2}$ is then determined
to be $m_{T2}^{\rm max} = 232.6\pm 1.5$ GeV for $\tilde m_\nu=80$
GeV. The $m_{T2}^{\rm max}$ as a function of $\tilde m_\nu$ is shown
in Fig. \ref{fig:mt2function}(b). Fitting the data points to the
theoretical curve (\ref{curve}) considering $m_t$ as a free
parameter while  using $m_W=80.45$ GeV and $m_b=4.7$ GeV, we obtain
\begin{eqnarray}
m_t = 170.5 \pm 0.5~ {\rm GeV},
\end{eqnarray}
which is quite close to the input top quark mass  $m_t=170.9$ GeV.
The uncertainty due to a variation of $m_b$ is negligible as it is
of ${\cal O}(m_b\delta m_b/m_t)$.  To check the effect of the $W$
boson mass, we repeated the fitting procedure while varying  $m_W$
by $\pm 0.5$ GeV.
The resulting shift of  $m_t$ turns out to be negligible.

\subsection{Template binned likelihood fit}
\label{template}

Perhaps the most reliable way to determine $m_t$ using $m_{T2}$ is
to employ the template binned likelihood fit. For this,  we
attempted to fit the $m_{T2}$ distribution of the `nominal data'
(which was generated with $m_t=170.9$ GeV) to `templates'. Here, a
template means a simulated $m_{T2}$ distribution with an input top
quark mass different from 170.9 GeV. The templates were generated
with input top quark mass between 166 GeV and 176 GeV, in steps of
1 or 0.5 GeV, using the same PYTHIA+PGS Monte Carlo programs as the case
of nominal data sample.

\begin{figure}[ht!]
\vskip 0.6cm
\begin{center}
\epsfig{figure=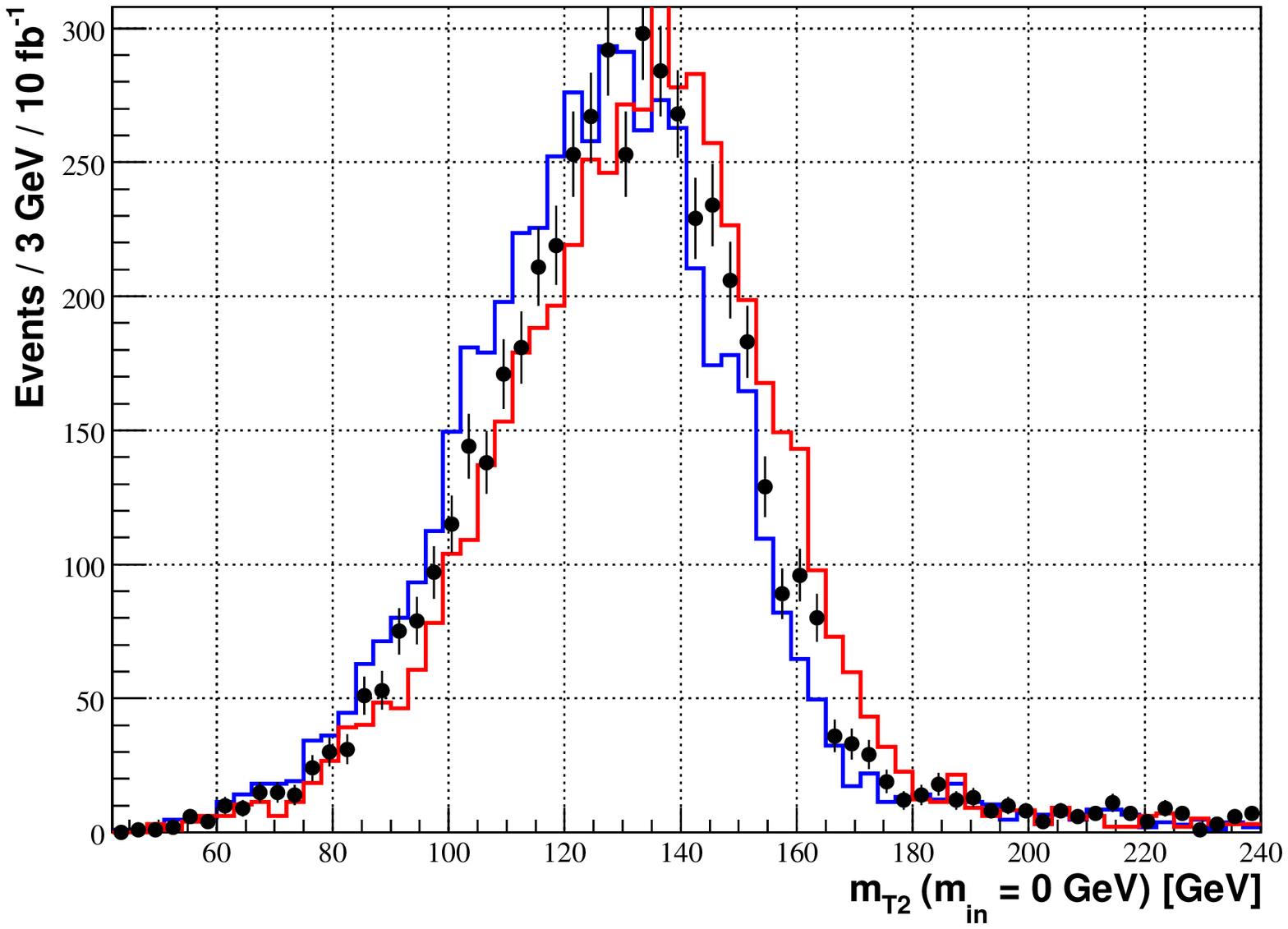,width=7cm,height=7cm}
\epsfig{figure=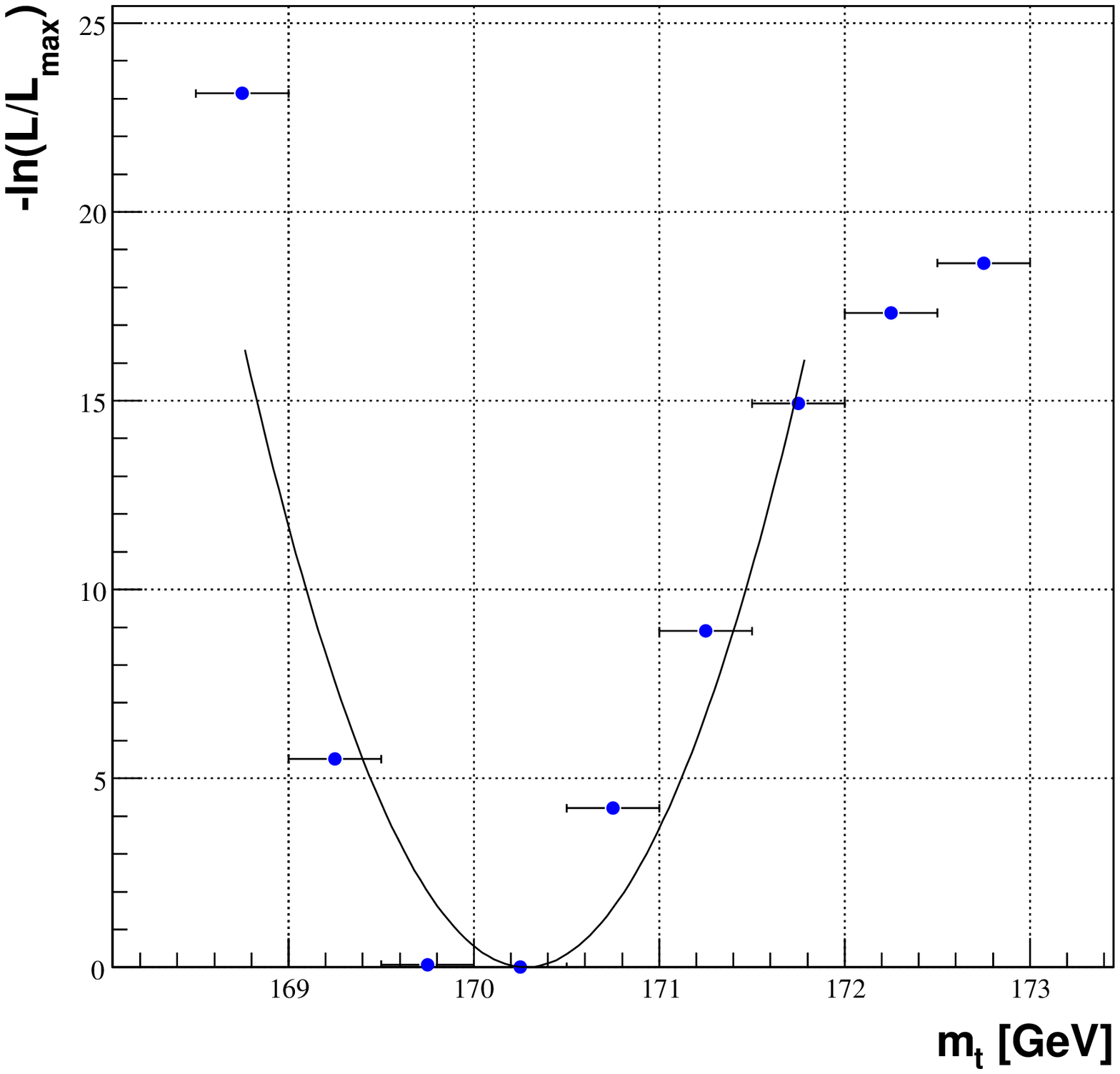,width=7cm,height=7cm}
\end{center}
\vskip 0.4cm \caption{\it (a) Three representative $m_{T2}$
distributions for the nominal data (points) and two templates with
$m_t=166$ GeV (blue solid) and $m_t=176$ GeV (red solid),
respectively. (b) The negative logarithm of the likelihood ratio
${{\cal L}/{\cal L}_{\rm max}}$ as a function of $m_t$ for the
$m_{T2}$ fit.} \label{fig:templates}
\end{figure}

Fig. \ref{fig:templates}(a) shows three representative $m_{T2}$
distributions for the nominal data (points) and two templates with
$m_t=166$ GeV (blue solid) and $m_t=176$ GeV (red solid),
respectively. Each template distribution is normalized to make the
total number of events is same as that of the nominal data. One can
notice that those three $m_{T2}$ distributions are well separated
from each other, showing the sensitivity of the $m_{T2}$
distribution to the input top quark mass.

Each template distribution is compared to the nominal data
distribution for a calculation of the logarithm of the binned
likelihood. The binned likelihood is defined as the product of the
Poisson probability for each bin over the $N$ bins in the fit range:
\begin{eqnarray}
{\cal L} = \prod_{i=1}^{N} {e^{-m_i} m_i^{n_i} \over n_i !},
\end{eqnarray}
where $n_i$ and $m_i$ are the event numbers at the $i$-th bin in the
distributions of the nominal data and the normalized template,
respectively.
The minimum of $-{\rm ln} {\cal L}$ gives the best fit value of the
top quark mass. We have chosen the $1\sigma$ deviated value of the
top quark mass as the one increasing $-{\rm ln} {\cal L}$ by 1/2.

We fit the $m_{T2}$ distribution of nominal data to templates
in the range 100 GeV $< m_{T2} <$ 180 GeV.
The result of the likelihood fit for $m_{T2}$ distributions is shown
in Fig. \ref{fig:templates}(b), where the negative logarithm of the
likelihood ratio ${\cal L}/{\cal L}_{\rm max}$ as a function of
$m_t$ is depicted. The ${\cal L}_{\rm max}$ is the maximum
likelihood which was determined as the minimum of a parabola fit to
the $-{\rm ln} {\cal L}$ distribution. The top quark mass resulting
from our template likelihood fit is given by
\begin{eqnarray}
m_t = 170.3 \pm 0.3 ~{\rm GeV},
\end{eqnarray}
which reproduces well the input top quark mass with a small
statistical error.

Although a detailed analysis of systematic uncertainties in the
template fit method is beyond the scope of this work, we expect that
systematic errors from b-jet energy scale, ISR/FSR and PDF are
also at the level of 1 GeV as those in the endpoint fit method
discussed in subsection \ref{endfit}.




%
%

\section{Conclusion}

We have examined the possibility to determine the top quark mass
using the $m_{T2}$ distribution of the dileptonic decay channel of
$t\bar{t}$ events at the LHC. For this, we have performed  three
Monte Carlo studies for the events produced at the LHC with 10
$fb^{-1}$ integrated luminosity: the first to fit the $m_{T2}$
distribution near the end point (for the neutrino mass $m_\nu=0$)
with an empirical function, the second to fit the functional
dependence of $m_{T2}^{\rm max}$ on the trial neutrino mass
$\tilde{m}_\nu\neq 0$, and the third to perform a template binned
likelihood fitting. It is found that the top quark mass can be
determined by the $m_{T2}$ variable alone
with a good precision at the level of 1 GeV.

\begin{acknowledgments}
We would like to thank T. Kamon and J. Lykken for asking about
the possibility to determine $m_t$ using $m_{T2}$ variable, and 
K. Kong for useful discussion.
This work was supported by the Korea Research Foundation Grant
funded by the Korean Government (MOEHRD, Basic Research Promotion
Fund) (KRF-2005-210-C000006), the Center for High Energy Physics of
Kyungpook National University, and the BK21 program of Ministry of
Education.

\end{acknowledgments}


\begin{thebibliography}{99}

\bibitem{atlas}
ATLAS Technical Proposal, CERN-LHCC-94-43.

\bibitem{cms}
CMS Physics Technical Design Report, CERN-LHCC-2006-021.

\bibitem{ttnlo}
R. Bonciani et al., Nucl. Phys. {\bf B 529} (1998) 424.

\bibitem{atlas-top}
I. Borjanovic et al., Eur. Phys. J. {\bf C39S2} (2005) 63-90 [hep-ex/0403021].

\bibitem{lester}
C.G.Lester and D.J.Summers, Phys. Lett. {\bf B 463} (1999) 99;
A.Barr, C.Lester, and P.Stephens, J. Phys. {\bf G 29} (2003) 2343.

\bibitem{cho}
W.S.Cho, K.Choi, Y.G.Kim and C.B.Park, to appear at Phys. Rev. Lett.
[arXiv:0709.0288]; W.S.Cho, K.Choi, Y.G.Kim and C.B.Park, JHEP 0802
(2008) 035 [arXiv:0711.4526].

\bibitem{ben}
B. Gripaios, JHEP {\bf 0802} (2008) 053 [arXiv:0709.2740].

\bibitem{lester2}
A.J.Barr, B.Gripaios and C.G.Lester, JHEP {\bf 0802} (2008) 014
[arXiv:0711.4008].

\bibitem{others}
G. Ross and M. Serna, arXiv:0712.0943 [hep-ph];
M. Nojiri, Y. Shimizu, S. Okada and K. Kawagoe, arXiv:0802.2412 [hep-ph].

\bibitem{pythia}
 T. Sjostrand, P. Eden, C. Friberg, L. Lonnblad, G. Miu, S. Mrenna and
 E. Norrbin, Computer Physics Commun. 135 (2001) 238;
 T. Sjostrand, S. Mrenna and P. Skands,
 LU TP 06-13, FERMILAB-PUB-06-052-CD-T [hep-ph/0603175].

\bibitem{pdf}
H. L. Lai {\it et al.} [CTEQ Collaboration], Eur. Phys. J. C {\bf 12} (2000) 375
[arXiv:hep-ph/9903282].

\bibitem{pgs}
http://www.physics.ucdavis.edu/$\sim$conway/research/software
/pgs/pgs4-general.htm.

\bibitem{wiki}
http://v1.jthaler.net/olympicswiki/doku.php.

\bibitem{alpgen}
M.L. Mangano, M. Moretti, F. Piccinini, R. Pittau and A.D. Polosa,
JHEP {\bf 0307} (2003) 001.

\bibitem{acermc}
B.P. Kersevan and E. Richter-Was, hep-ph/0405247;
Comput. Phys. Commun. {\bf 149} (2003) 142.

\end{thebibliography}
\end{document}